\documentclass[sigconf]{acmart}
\usepackage{flushend}
\usepackage{url}

\usepackage{diagbox}
\usepackage{array}
\usepackage{graphicx}
\usepackage{adjustbox}

\AtBeginDocument{%
  \providecommand\BibTeX{{%
    \normalfont B\kern-0.5em{\scshape i\kern-0.25em b}\kern-0.8em\TeX}}}

\setcopyright{acmcopyright}
\copyrightyear{2024}
\acmYear{2024}

\acmBooktitle{ICCA ’24, September 17–18, 2024}

\begin{document}

\title[Evaluating Large Language Models]{Programming with AI: Evaluating ChatGPT, Gemini, AlphaCode, and GitHub Copilot for Programmers}


\author{Md Kamrul Siam}
\affiliation{%
	\institution{New York Institute of Technology}
	\streetaddress{1 Th{\o}rv{\"a}ld Circle}
	\city{New York}
        \state{NY}
	\country{USA}}
\email{ksiam01@nyit.edu}

\author{Huanying Gu}
\affiliation{%
	\institution{New York Institute of Technology}
	\streetaddress{1 Th{\o}rv{\"a}ld Circle}
	\city{New York}
        \state{NY}
	\country{USA}}
\email{hgu03@nyit.edu}

\author{Jerry Q. Cheng}
\affiliation{%
	\institution{New York Institute of Technology}
	\streetaddress{1 Th{\o}rv{\"a}ld Circle}
	\city{New York}
        \state{NY}
	\country{USA}}
\email{jcheng18@nyit.edu}

\renewcommand{\shortauthors}{Md Kamrul Siam et al.}
\renewcommand{\shortauthors}{Manuscript Submitted to ACM}

\begin{abstract}

 
Our everyday lives now heavily rely on artificial intelligence (AI) powered large language models (LLMs). Like regular users, programmers are also benefiting from the newest large language models. In response to the critical role that AI models play in modern software development, this study presents a thorough evaluation of leading programming assistants, including ChatGPT, Gemini (Bard AI), AlphaCode, and GitHub Copilot. The evaluation is based on tasks like natural language processing and code generation accuracy in different programming languages like Java, Python and C++. Based on the results, it has emphasized their strengths and weaknesses and the importance of further modifications to increase the reliability and accuracy of the latest popular models. Although these AI assistants illustrate a high level of progress in language understanding and code generation, along with ethical considerations and responsible usage, they provoke a necessity for discussion. With time, developing more refined AI technology is essential for achieving advanced solutions in various fields, especially with the knowledge of the feature intricacies of these models and their implications. This study offers a comparison of different LLMs and provides essential feedback on the rapidly changing area of AI models. It also emphasizes the need for ethical developmental practices to actualize AI models’ full potential.
  
\end{abstract}






\maketitle

\section{Introduction}
The advent of the AI concept presents a new revolutionary age of innovation with an AI model and LLM-powered chatbots changing how our software is being developed and problems solved ~\cite{gates_age_nodate}. With the launch of ChatGPT and the newest LLM tools, such as GitHub Copilot, Bard AI (which is now a part of the Gemini framework), and DeepMind's AlphaCode, which have been developed by major players in the industry like Google, GitHub, and OpenAI, these AI systems have captured the attention of the tech community with their capacity to understand languages and generate programming languages. The reality of AI assistants is that they are revolutionary and keep widening the limits of AI models at work. Therefore, the discussion about their accuracy, architecture, capabilities, and implications for the future of AI technologies is crucial. One of the first and most effective LLMs, ChatGPT, attracted 100 million users in just two months after the launch, making it the fastest-growing platform out of all those based on technology and a testament to how much the consumers needed such platforms ~\cite{milmo_chatgpt_2023}. With the help of LLMs, notable progress in code generation and Natural language Processing (NLP) has been made recently~\cite{noauthor_what_2024}. One example is the generative pre-trained transformer (GPT) model series\cite{Radford_2018}. These models, which have received extensive training on textual data show that they can produce codes on the same level as human written codes and execute language-based tasks with remarkable accuracy. This paper thoroughly studies the most recent large language models, highlighting their strengths and weaknesses and crucially contributing to responsible development practices in the benefits of AI models in various fields. Thus, to understand and capture the behaviour of popular LLMs, we pose three research questions:

\begin{itemize}

    \item \textbf{RQ1:} Which model provides the most accurate code for programmers?
    \item \textbf{RQ2:} What are the metrics are frequently used to evaluate LLM generated codes?
    \item \textbf{RQ3:} What are the benchmarks are being used to evaluate LLM generated codes?
    
\end{itemize}

\section{RELATED WORK}

The period of language models started in the late 19th century with the development of mathematical models known as Statistical Language Models (SLMs), which provide a probabilistic statistical framework for handling contextually important aspects of natural language~\cite{jelinek_statistical_1997}. SLM was among the pioneer techniques. After SLMs were developed, neural network-based machine learning (NLM) entered the revolution and predicted the likelihood of words in a sequence~\cite{kombrink11_interspeech, NIPS2000_728f206c}. PLM (Pre-trained Language Model) was the most recent language model prior to the development of more recent LLMs like chatGPT, Genimi, GitHub Copilot and AlphaCode. PLMs have a significant impact on the LLM industry, which came about after the introduction of NLM~\cite{devlin-etal-2019-bert, noauthor_plm_2023}. Pre-training is the first phase of training that PLMs go through with a large amount of unlabeled text to help them understand basic language structures including vocabulary, syntax, semantics, and logic. Research on generative models—which are trained by gathering a large amount of data in a particular area, such sounds, phrases, or images—was published by OpenAI on June 16, 2016. The model is then taught to produce comparable data and research on optimizing the GPT-2 language model with human preferences and input was published by OpenAI on September 19, 2019~\cite{hines_history_2023}. As part of a free research preview, OpenAI released ChatGPT utilizing GPT-3.5 on November 30, 2022. GitHub Copilot, a tool for code completion from GitHub and OpenAI, was announced on June 29, 2021, as being in the technical preview in the Visual Studio Code development environment~\cite{dohmke_github_2022, gershgorn_github_2021}. It was released eventually as a plugin to the JetBrains marketplace on October 29, 2021, and became available to all developers worldwide on June 21, 2022~\cite{noauthor_github_nodate}. As part of a free research preview, OpenAI released ChatGPT, a chatbot utilizing GPT-3.5 on November 30, 2022. This chatbot is built on top of large language models (LLMs). It lets users adjust and tailor a conversation to their preference regarding length, format, style, level of detail, and language. By January 2023, it had registered over 100 million users~\cite{milmo_chatgpt_2023}. Google unveiled BARD AI, a generative artificial intelligence chatbot, to the globe on February 6, 2023~\cite{dastin_google_2023}. It commenced limited operations in March 2023, and the product was available in more regions by May. In February 2024, Bard merged under the banner of Gemini~\cite{dastin_google_2024}.
In contrast, in 2022, Goolgle's DeepMind introduced AlphaCode. This new AI-powered code writing program can create computer programs at the level of a professional programmer, using the system to compete in Codeforces, coding challenges popular with the programming community, and managing to rank 54\% in the median Codeforces score after being trained on GitHub data and Codeforces problems and solutions~\cite{noauthor_competitive_2022}. The task was set to make up a particular answer that was unique and different.

\section{TRANSFORMER ARCHITECTURE}

The Transformer architecture seems indispensable for programming and central to the Large Language Models (LLMs)~\cite{noauthor_understanding_2024, noauthor_gpt-35_2023}. This way, it is tuned for tasks such as machine translation or inducing general-purpose text generation. In a nutshell, the Transformer utilizes the mechanism of self-attention to process input data. Unlike the sequence data models (RNNs and LSTM) employed for natural language processing in previous deep learning models, Transformers aim to process parallel input data. Apart from this, the model's efficiency is also heightened, and there is an improved capability to understand the context within language~\cite{perdigao_chatgpt_2023}. 

Transformer architecture is the main idea behind a few AI models, such as AlphaCode, GitHub Copilot, ChatGPT, and BardAI (laying a foundation for the upcoming Genie model). Every model innovates by using the transformer architecture differently to suit its purpose and the training dataset on which it is trained. Transformers’ attention technique, which enables them to process and grasp intricate language patterns efficiently, is the primary advantage of this model. Via a process in which only the significant pieces are acknowledged, the models generate responses with more conciseness and pertinence. The model uses an extensive training corpus to grasp language subtleties and give discrete outputs. The transformer algorithm, considered a landmark in NLP, has shown a new way to develop chatting AI to participate in human-like conversations. Through this, the applications of transformer architectures in machine learning are proven. 

DeepMind's AlphaCode exemplifies the significance of transformers as a force multiplier. Within this framework, a code generator model, designed to generate competitive programming problems related to codes, utilizes an encoder-decoder transformer model. It employs an additional reduced competitive programming dataset and uses large-scale model sampling to explore the space~\cite{li_competition-level_2022}. This model is performance-sensitive; it can be reduced to some selected submissions, demonstrating the effectiveness of the transformer-based architectures~\cite{dibia_alphacode_2022}. Codex is the system used to build GitHub Copilot; an application based on transformer architectures. This model has been implemented on large numbers of code repositories. Similarly to language models such as GPT-3 and LaMDA, the architecture of transformers works well with sequential data, be it text, lines of software code, or amino acid sequences~\cite{chandra_transformer-based_2023, noauthor_transformer_nodate}. A transformer architecture can be applied to many tasks, including identifying the following words in a sentence or prior computer instructions, proving transformer models' effectiveness. 

\begin{figure*}
    \centering
    \includegraphics[width=0.6\paperwidth,height=200pt]{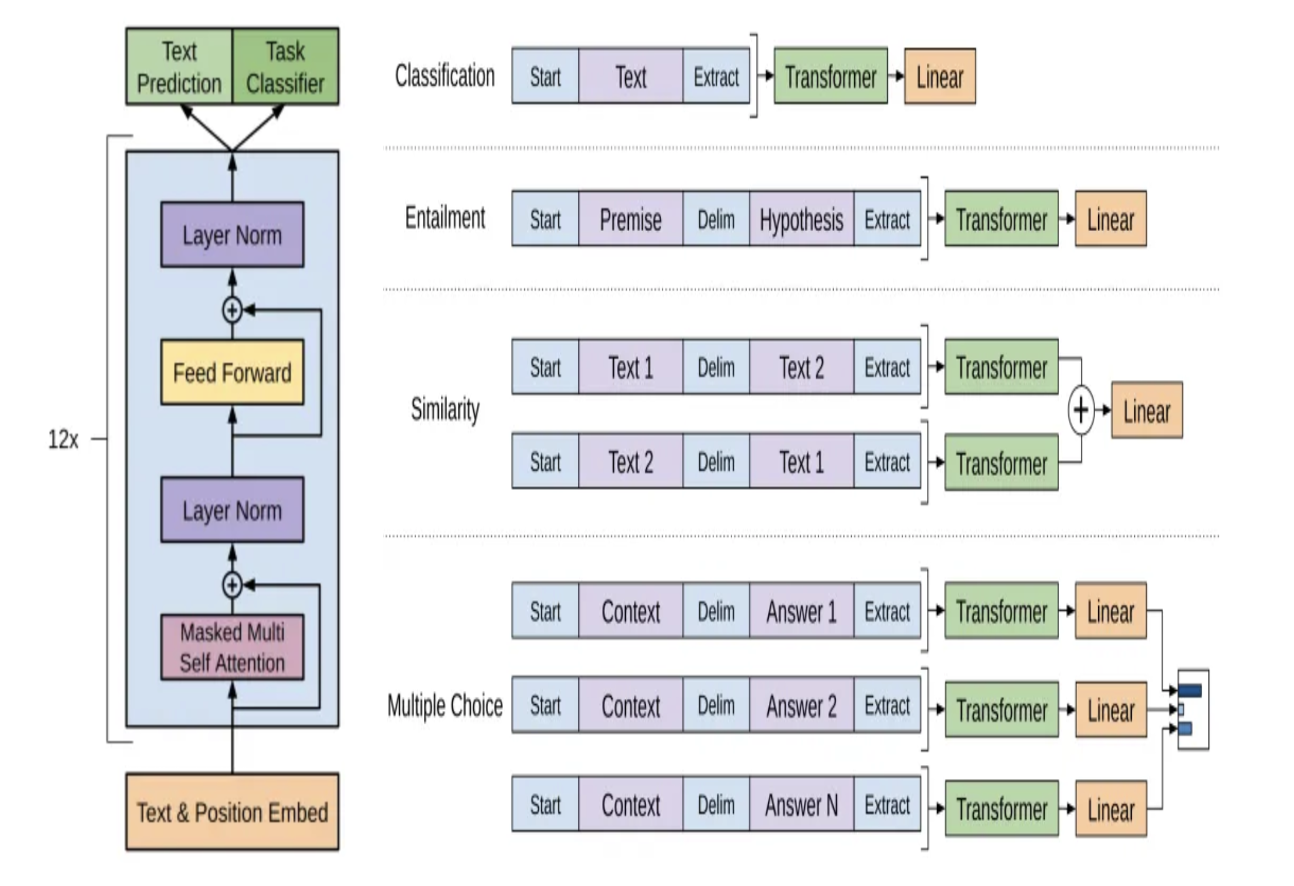}
    \caption{ChatGPT Architecture~\cite{noauthor_gpt-35_2023}}
    \label{fig:enter-label}
\end{figure*}

ChatGPT, a GPT model variant with the exact underlying mechanism, is another example of a transformer-based neural network ~\cite{roumeliotis_chatgpt_2023}. It consists of the encoder and the decoder, similar to the traditional transformer. The encoder, composed of several layers of self-attention and feed-forward neural networks, processes and interprets the input text ~\cite{xiao_introduction_2023}. The decoder, also made of self-attention-based and feed-forward neural network layers, generates the output text, demonstrating the greatness of transformer architecture in text generation~\cite{phd_decoder-only_2024, noauthor_gpt-35_2023, xiao_introduction_2023}. 


Bard AI, now known as Gemini, is another model resembling the Transformer architecture~\cite{kulkarni_google_2023}. It is trained on the linguistic statistics between the words and phrases as a vast corpus of text~\cite{santosh_large-language-models_2024}. With the tools to understand an extensive repertoire of text and code, Bard can create human-like text, translate languages, write various types of creative work, and provide meaningful responses~\cite{noauthor_code_2023}. 

\subsection{ChatGPT}
ChatGPT is a huge language model-based chatterbot that was developed by the OpenAI organization in November 2022. It also could be applied to the list of tasks, such as writing, analysis, and solving problems. The new player on the street, ChatGPT, is the talk of the town due to its unique aptitude of writing things in a similar way like a human being does. On the other hand, while some individuals might misuse this technology and it may not give correct outputs to us~\cite{mollick_chatgpt_2022}. With ChatGPT and other similar AI virtual assistants becoming more mature and common, the ethical aspect of the matter as well as the necessity for duty-bound development and the deployment of the technology has received wide attention in the current discussion.

\subsection{Gemini}
Gemini is a multimodal AI meaning it has the ability to comprehend and write content using various data modalities, such as text, images, audio, video, and code. Perhaps the most spectacular thing is that it suits almost any feature. It is worth talking about the work of Gemini on the MMLU task. MMLU is a short test that seeks to determine an AI system's language processing competence and sentence construction skills for various tasks. The effectiveness of Gemini, therefore, can be boiled down to the superb language use and problem-solving faculties it is endowed with~\cite{noauthor_gemini_nodate}. Thanks to the Gemini’s API, developers can design their apps of around Gemini functionality. It is a multi-lingual software with support for languages like Python, Go, Node.js, and Swift. As a result, it has boosted the chances of getting popular among other developers as well. Finally, Gemini offers three model sizes: 1.0 Ultra, 1.0 Pro, and 1.0 Nano. These opportunities allow developers to select the best model for their needs~\cite{noauthor_build_nodate}.

\subsection{AlphaCode}

The new approach of code creation that has been developed is known as AlphaCodium, and it was created by Tal Ridnik~\cite{ridnik_state---art_2024}. This is a multistage, process-based, code-oriented iterative approach that guarantees improved code problem performance for LLMs. Consideration has been given to the types of AI and their characteristics for code analysis in software design and development tasks, including code and document generation3. Three areas of skills are required for machine learning models to handle structured expression (SE) tasks involving code analysis: syntax, static behavior comprehension, and dynamic behavior comprehension. 

We already have some existing research those were designed to make it easier for LLMs to understand and interpret abstract code structures like AST, CFG, and CG. Four cutting-edge foundational models were used: CodeLlama-13b-instruct, GPT4, GPT 3.5, and StarCoder~\cite{ma_lms_2023}. One of the topics explored in the research was the ability of LLMs to analyze code syntax; nevertheless, they are unable to comprehend code semantics, including dynamic semantics. AlphaCode would be helpful in order to support those study findings and needs for programmers and software developers in their daily tasks.

Although LLMs are capable of creating facts and making sense out of code structures, they are not real. These results imply that more research is necessary to create a system for guaranteeing the validity of LLM results. Because AlphaCodium is a multi-stage, iterative, test-driven process that focuses on code, code problems, in particular, improve LLM performance~\cite{ridnik_code_2024}. The CodeContests dataset, which consists of competitive programming issues from online platforms like Codeforces, was used to highlight this approach. Better results are being produced in a noticeable and prominent manner by the projected flow. As a result, GPT-4’s accuracy(pass@5) improved dramatically from 19\% to 44\% due to the introduction of an AlphaCodium flow, as compared to employing a simple, lengthy prompt~\cite{ridnik_code_2024}. The majority of methods used in this undertaking and the overall abstraction of the peace plan are applicable to a variety of coding tasks.

\subsection{GitHub Copilot}

GitHub Copilot is an AI-based code recommendation tool designed by GitHub. It acts as an AI code pair partner, helping you complete your code by giving suggestions as you type. It gives you the context of your code by auto-completing lines and functions. By doing so, you will recognize alternate ways of solving problems, write tests, and discover new APIs~\cite{friedman_introducing_2021}. It can be installed as a plugin into your favored platform. It can be done using an individual subscription to GitHub Copilot or an organization subscription to GitHub Copilot Business~\cite{noauthor_quickstart_nodate, noauthor_about_nodate}. Filters are present in GitHub Copilot that eliminate offensive words in the prompts and generate phrases that are not sensitive. As a rule, GitHub Copilot is a tool that utilizes AI to enhance the coding experience by providing helpful hints and advice directly to an editor.

\section{Methods}

In order to support the research questions, we analyzed 10 latest research papers that were published between 2022 and 2024~\cite{liu_refining_2024,Akter_Yu_Muhamed_Ou_Bäuerle_Cabrera_Dholakia_Xiong_Neubig_2023,OpenAI_Achiam_Adler_Agarwal_Ahmad_Akkaya_Aleman_Almeida_Altenschmidt_Altman_et_al_2023,Zhang_Zhao_Liu_Zheng_Qi_Gu_Zhang_Dong_Tang_2024,Our_next_generation_model_2024, Gemini_Team_Anil_Borgeaud_Alayrac_Yu_Soricut_Schalkwyk_Dai_Hauth_Millican_et_al_2023,Paul_Zhu_Bayley_2024,Li_Choi_Chung_Kushman,Siroš_Singelée_Preneel_2024, Nguyen_Nadi_2022}. We further attempted to determine which model performed based on two major metrics: pass@k and Test Case Pass Rate, which are described in Table 1. Furthermore, we conducted a comparison analysis using those evaluation metrics. 

\textbf{Pass@k}: This metric assesses the likelihood that at least one of the k produced code samples passes all test cases for a particular problem. For example, the measure pass@1 is widely used across several models, including ChatGPT, Gemini, GitHub Copilot and AlphaCode to indicate the accuracy rate when only one generated sample is examined. Higher k values, such as pass@100, are used in some evaluations (e.g., Gemini-Ultra), allowing the model more attempts and offering a more comprehensive assessment of the model's capabilities when given more opportunities to create accurate solutions.

\textbf{Test Case Pass Rate}: This metric measures the percentage of test cases that the generated code successfully passes. It is mostly used to evaluate AlphaCode and GitHub Copilot on platforms such as Codeforces and LeetCode. This statistic provides a detailed evaluation of the model's capacity to generate functionally valid code across a wide range of test situations, reflecting its accuracy and resilience.

\begin{table}[h!]
    \centering
    \begin{adjustbox}{max width=\linewidth}
    \begin{tabular}{|c|c|c|c|c|c|c|}
        \hline
        \textbf{Model} & \textbf{Year} & \textbf{Metrics} & \textbf{Evaluation Benchmark / Standard} & \textbf{Programming language} &\textbf{Accuracy} & \textbf{Reference} \\ 
        \hline
        ChatGPT & 2023 & pass@1 & LeetCode (easy, medium, and hard) & Python & 0.664 & ~\cite{liu_refining_2024}\\
        \hline
        ChatGPT & 2023 & pass@1 & LeetCode (easy, medium, and hard) & Java & 0.691 & ~\cite{liu_refining_2024}\\
        \hline
        
       Gemini Pro & 2023 & pass@1 & HumanEval & Python & 0.598 & ~\cite{Akter_Yu_Muhamed_Ou_Bäuerle_Cabrera_Dholakia_Xiong_Neubig_2023}\\
        \hline
       Gemini Pro & 2023 & pass@1 & ODEX & Python & 0.399 & ~\cite{Akter_Yu_Muhamed_Ou_Bäuerle_Cabrera_Dholakia_Xiong_Neubig_2023}\\
        \hline

         ChatGPT (3.5 Turbo) & 2023 & pass@1 & HumanEval & Python & 0.743 & ~\cite{Akter_Yu_Muhamed_Ou_Bäuerle_Cabrera_Dholakia_Xiong_Neubig_2023}\\
        \hline
        ChatGPT (3.5 Turbo) & 2023 & pass@1 & ODEX & Python & 0.526 & ~\cite{Akter_Yu_Muhamed_Ou_Bäuerle_Cabrera_Dholakia_Xiong_Neubig_2023}\\
        \hline

         ChatGPT (4 Turbo) & 2023 & pass@1 & HumanEval & Python & 0.768  & ~\cite{Akter_Yu_Muhamed_Ou_Bäuerle_Cabrera_Dholakia_Xiong_Neubig_2023}\\
         
        \hline

         ChatGPT (4 Turbo) & 2023 & pass@1 & ODEX & Python & 0.458  & ~\cite{Akter_Yu_Muhamed_Ou_Bäuerle_Cabrera_Dholakia_Xiong_Neubig_2023}\\
         
        \hline

       ChatGPT (GPT 4) & 2023 & pass@1 & Natural Code Bench & Java and Python & 0.528  & ~\cite{OpenAI_Achiam_Adler_Agarwal_Ahmad_Akkaya_Aleman_Almeida_Altenschmidt_Altman_et_al_2023,Zhang_Zhao_Liu_Zheng_Qi_Gu_Zhang_Dong_Tang_2024}\\
       \hline
        
       ChatGPT (GPT-4) & 2023 & pass@1 & HumanEval & Java and Python & 0.805  & ~\cite{OpenAI_Achiam_Adler_Agarwal_Ahmad_Akkaya_Aleman_Almeida_Altenschmidt_Altman_et_al_2023,Zhang_Zhao_Liu_Zheng_Qi_Gu_Zhang_Dong_Tang_2024}\\

       \hline
        
       ChatGPT (GPT-4-Turbo-0125) & 2023 & pass@1 & Natural Code Bench  & Java and Python & 0.525  & ~\cite{OpenAI_Achiam_Adler_Agarwal_Ahmad_Akkaya_Aleman_Almeida_Altenschmidt_Altman_et_al_2023,Zhang_Zhao_Liu_Zheng_Qi_Gu_Zhang_Dong_Tang_2024}\\

       \hline
        
       ChatGPT (GPT-4-Turbo-0125) & 2023 & pass@1 &HumanEval  & Java and Python & 0.872  & ~\cite{OpenAI_Achiam_Adler_Agarwal_Ahmad_Akkaya_Aleman_Almeida_Altenschmidt_Altman_et_al_2023,Zhang_Zhao_Liu_Zheng_Qi_Gu_Zhang_Dong_Tang_2024}\\

       \hline
        
       ChatGPT (GPT-4-Turbo-1106) & 2023 & pass@1 & Natural Code Bench  & Java and Python & 0.515  & ~\cite{OpenAI_Achiam_Adler_Agarwal_Ahmad_Akkaya_Aleman_Almeida_Altenschmidt_Altman_et_al_2023,Zhang_Zhao_Liu_Zheng_Qi_Gu_Zhang_Dong_Tang_2024}\\

        \hline
        
       ChatGPT (GPT-4-Turbo-1106) & 2023 & pass@1 & HumanEval & Java and Python & 0.817  & ~\cite{OpenAI_Achiam_Adler_Agarwal_Ahmad_Akkaya_Aleman_Almeida_Altenschmidt_Altman_et_al_2023,Zhang_Zhao_Liu_Zheng_Qi_Gu_Zhang_Dong_Tang_2024}\\

       \hline
        
       ChatGPT (GPT-3.5-Turbo) & 2023 & pass@1 & Natural Code Bench  & Java and Python & 0.407  & ~\cite{OpenAI_Achiam_Adler_Agarwal_Ahmad_Akkaya_Aleman_Almeida_Altenschmidt_Altman_et_al_2023,Zhang_Zhao_Liu_Zheng_Qi_Gu_Zhang_Dong_Tang_2024}\\

       \hline
        
       ChatGPT (GPT-3.5-Turbo) & 2023 & pass@1 & HumanEval  & Java and Python & 0.652  & ~\cite{OpenAI_Achiam_Adler_Agarwal_Ahmad_Akkaya_Aleman_Almeida_Altenschmidt_Altman_et_al_2023,Zhang_Zhao_Liu_Zheng_Qi_Gu_Zhang_Dong_Tang_2024}\\
       
         \hline
        
       Gemini-1.5-Pro & 2024 & pass@1 & Natural Code Bench  & Java and Python & 0.423  & ~\cite{Our_next_generation_model_2024,Zhang_Zhao_Liu_Zheng_Qi_Gu_Zhang_Dong_Tang_2024}\\

       \hline
        
       Gemini-1.5-Pro & 2024 & pass@1 & HumanEval & Java and Python & 0.719  & ~\cite{Our_next_generation_model_2024,Zhang_Zhao_Liu_Zheng_Qi_Gu_Zhang_Dong_Tang_2024}\\

       \hline
        
       Gemini-Ultra (Transformer) & 2023 & pass@100 & HumanEval & Python & 0.747  & ~\cite{Paul_Zhu_Bayley_2024, Gemini_Team_Anil_Borgeaud_Alayrac_Yu_Soricut_Schalkwyk_Dai_Hauth_Millican_et_al_2023}\\

       \hline
        
       Gemini-Ultra (Transformer) & 2023 & pass@100 & Natural2Code & Python & 0.749  & ~\cite{Paul_Zhu_Bayley_2024, Gemini_Team_Anil_Borgeaud_Alayrac_Yu_Soricut_Schalkwyk_Dai_Hauth_Millican_et_al_2023}\\

        \hline
        
      Gemini-Pro (Transformer) & 2023 & pass@100 & HumanEval & Python & 0.677  & ~\cite{Paul_Zhu_Bayley_2024, Gemini_Team_Anil_Borgeaud_Alayrac_Yu_Soricut_Schalkwyk_Dai_Hauth_Millican_et_al_2023}\\

        \hline
        
      Gemini-Pro (Transformer) & 2023 & pass@100 & Natural2Code & Python & 0.696  & ~\cite{Paul_Zhu_Bayley_2024, Gemini_Team_Anil_Borgeaud_Alayrac_Yu_Soricut_Schalkwyk_Dai_Hauth_Millican_et_al_2023}\\
       
        \hline
       AlphaCode & 2022 & Test Case (Codeforces) & Test case pass rate & C++ & 0.45  & ~\cite{Li_Choi_Chung_Kushman}\\

     \hline
     AlphaCode & 2022 & Test Case (Codeforces) & Test case pass rate & Python & 0.54  & ~\cite{Li_Choi_Chung_Kushman}\\
       
        \hline

       AlphaCode & 2022 & Test Case (Codeforces) & Test case pass rate & Java & 0.51  & ~\cite{Li_Choi_Chung_Kushman}\\

       \hline

       GitHub Copilot & 2024 & Test Case (LeetCode) & Test case pass rate & Java & 0.757  & ~\cite{Siroš_Singelée_Preneel_2024}\\

       \hline

       GitHub Copilot & 2024 & Test Case (LeetCode) & Test case pass rate & C++ & 0.733  & ~\cite{Siroš_Singelée_Preneel_2024}\\

         \hline

       GitHub Copilot & 2024 & Test Case (LeetCode) & Test case pass rate & Python3 & 0.669  & ~\cite{Siroš_Singelée_Preneel_2024}\\

        \hline

       GitHub Copilot & 2022 & Test Case (LeetCode) & Test case pass rate & Python & 0.42  & ~\cite{Nguyen_Nadi_2022}\\

       \hline

       GitHub Copilot & 2022 & Test Case (LeetCode) & Test case pass rate & Java & 0.57  & ~\cite{Nguyen_Nadi_2022}\\
       
        \hline
        
    \end{tabular}
    \end{adjustbox}
    \caption{Comparison of Code Generation Models Based on Evaluation Metrics and Benchmarks.}
    \label{tab:sample_table}
\end{table}

\section{Empirical Result}

\subsection{\textbf{RQ1:} Which model provides the most accurate code for programmers?}

\textbf{ChatGPT (GPT-4-Turbo-0125)} emerges as the model providing the most accurate code for programmers. It achieved a pass@1 accuracy of 87.2\% on the HumanEval benchmark. On the other hand, \textbf{ChatGPT (GPT-4-Turbo-1106)} has scored an accuracy of 81.7\% on the HumanEval benchmark, which are among the highest scores reported. These results indicate that \textbf{ChatGPT (GPT-4-Turbo-0125)} is highly effective in generating correct code on the first attempt, making it a top choice for developers who need reliable and precise code outputs.

Gemini-1.5-Pro also shows strong performance with a pass@1 accuracy of 74.9\% on the HumanEval benchmark. While slightly lower than ChatGPT, this still represents a high level of accuracy, making it another solid option for code generation.

For models evaluated with multiple attempts, Gemini-Ultra (Transformer) achieves a pass@100 accuracy of 74.7\% on Natural2Code, demonstrating that when allowed more attempts, this model also performs very well.

Overall, ChatGPT (GPT-4-Turbo-0125) stands out as the most accurate model for generating code across different benchmarks.

\subsection{\textbf{RQ2:} What are the metrics are frequently used to evaluate LLM generated codes?}

The most frequently used metrics to evaluate LLM-generated codes in the provided table are:

\textbf{Pass@k:} This is the most common metric used, indicating the success rate when only k number of code samples are generated and evaluated. It's used extensively across various models, including ChatGPT, Gemini, and others. We also discovered that 7 out of 10 the research papers used Pass@k metric.

\textbf{Test Case}: This metric is used mostly for models like AlphaCode and GitHub Copilot. It evaluates the percentage of test cases that the generated code passes, offering a detailed assessment of the model's performance in real-world coding scenarios. Our research supports that 3 out of 10 papers had the use of Test Case Metric.

These metrics help in comparing the effectiveness and reliability of different models in generating correct and functional code.

\subsection{\textbf{RQ3:} What are the benchmarks are being used to evaluate LLM generated codes?}

Table 2 demonstrates that 6 out of 10 research was conducted using HumanEval benchmark which is one of the widely used benchmark used to evaluate code generated by LLMs. Also, Test Case Pass Rate and Natural Code Bench are also becoming popular since HumanEval is a benchmark that has been using for a long time~\cite{liu_refining_2024}.

\begin{table}[h!]
\centering
\begin{adjustbox}{max width=\linewidth}
\begin{tabular}{|l|c|l|}
\hline
\textbf{Benchmark}            & \textbf{Papers} & \textbf{References} \\
\hline
HumanEval                     & 6               & [22], [54], [68], [21], [55], [63] \\
\hline
LeetCode                      & 1               & [48]           \\
\hline

Natural Code Bench            & 3               & [54], [68], [21]                       \\
\hline
ODEX                          & 1               & [22]                   \\
\hline
Natural2Code                          & 1               & [55], [63]                         \\
\hline
Test Case Pass Rate                          & 3               & [47], [62], [52]                         \\
\hline
\end{tabular}
\end{adjustbox}
\caption{Benchmarks Used to Evaluate LLM-Generated Codes and Their Usage in Academic Research.}
\label{tab:benchmarks}
\end{table}

\section{Context}

ChatGPT impresses with its ability to preserve discourse continuity by retrieving all the conversation material discussed. It relies on a bi-directional context window that is somewhat like real-time memory, enabling it to keep relevant tokens from the discourse~\cite{noauthor_models_nodate}. This, in effect, makes Communications with ChatGPT seem natural and yields human-like responses. On the other hand, the downside of the technique is the use of a context window that sometimes generates answers that sound plausible but might be incorrect. Gemini's comparable success to ChatGPT is mainly attributed to its ability to demonstrate contextual understanding aided by Google's resources and workforce and its flexible context window. Gemini then uses its data center with the potential to utilize diverse datasets and accurately carry out the computations, which contributes to the rising performance level.

Differences can be drawn between AlphaCode and GitHub Copilot in that they are primarily involved in programming-oriented operations. AlphaCode, created by DeepMind, is a system for code production with no boundaries that shows the capacity to find solutions to complex problems that involve learning from algorithms and natural language~\cite{noauthor_chatgpt_nodate}. It has demonstrated the ability to generate a code that can beat other programs in simulated competitions. It is already doing the middle work but that of complex solutions. By contrast, GitHub Copilot is an AI-backed tool for code completion that advises developers to make them guess the algorithm, finish the functions, and give context-based tips. Copilot applies a mixture of transformer-based classifiers and the contextual data diverted from the code editor to improve the developers' overall productivity and eliminate boilerplate code. However, these language models are developed with different spearheads, yet significant advances in AI-driven tools still occur in other spheres.
Search and Interactions

ChatGPT from OpenAI is a language model that uses Artificial Intelligence to create text similar to human language when considering context and past dialogues~\cite{cfa_how_2024}. It employs transformer-based language models, which allow it to generate the code at a scale never seen before~\cite{noauthor_what_nodate}. Gemini, from Google, is a soulmate for your creativity and productivity, and it aims to do so ~\cite{noauthor_gemini_nodate}. Google AI technology supports writing, planning, learning, and many other activities. GitHub Copilot is a code suggestion tool developed to improve code writing~\cite{noauthor_getting_nodate}. An AI double agent gives you a hint code as you are coding. You can use the power of GitHub Copilot by typing the code you need or by writing a natural language comment about what the code should perform~\cite{noauthor_about_nodate}

AlphaCode, the product from DeepMind, can write code in all programming languages for different tasks~\cite{li_competition-level_2022}. It transforms problem statements into code and accesses vast amounts of code to analyze and extract solutions from its patterns and learn new patterns~\cite{noauthor_chord_nodate}. Such technologies are compatible with many hosting environments and productive tools to maximize productivity and efficiency~\cite{basheer_unleashing_2023}. They use AI and machine learning to comprehend the context fully, create text that reads like human speech, give out code, and solve complicated problems. They are the known technological advancements in AI applications in chatting, coding, and problem-solving. Remember that these technologies might be powerful, but they are tools that should be applied cautiously to address a balanced approach~\cite{noauthor_gemini_2016}.

\section{RESPONSE ACCURACY}

One of the central metrics for LLMs is their response accuracy. This is the level of correctness, relevance, and coherence of their outputs~\cite{isci_comprehensive_2023}. The Google AI model Gemini has a built-in fact-checker feature that reached a median accuracy level of 3 when interpreting biochemical laboratory data~\cite{kaftan_response_2024}.

\begin{figure}
    \centering
    \includegraphics[width=250 pt,height=150 pt]{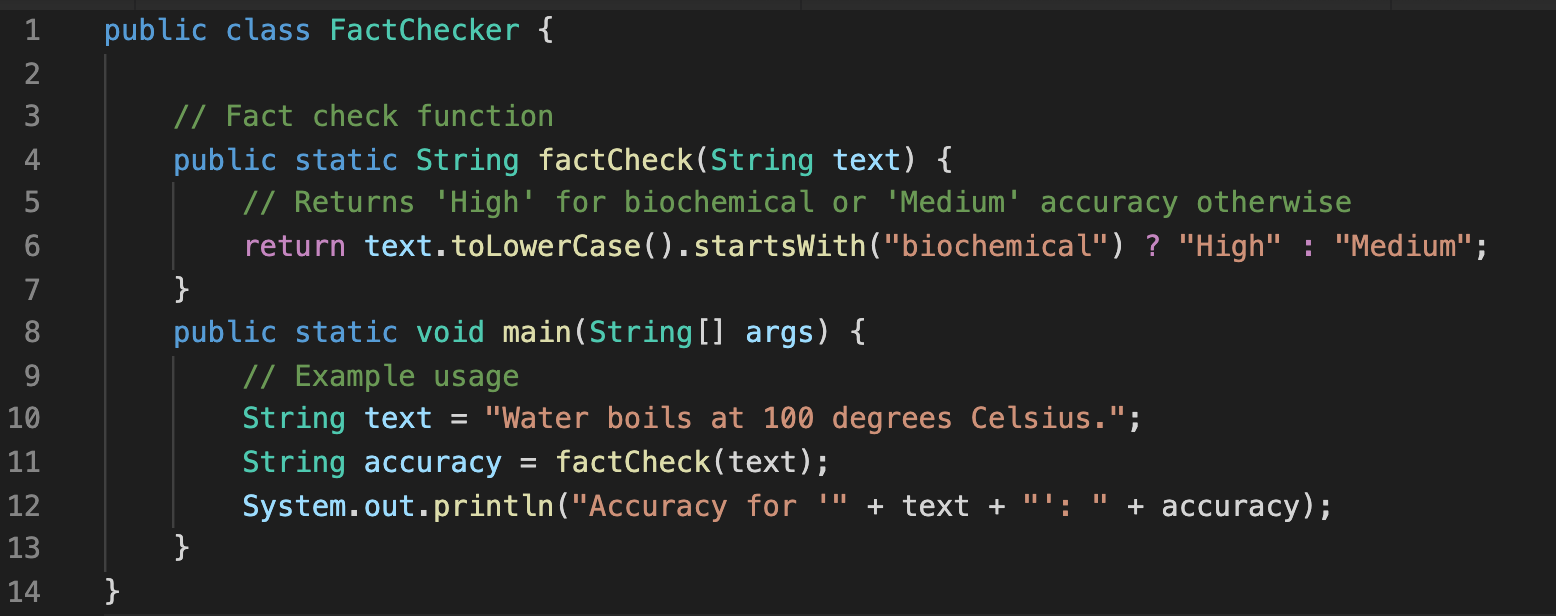}
    \caption{Code for checking fact using ChatGPT ~\cite{openai_chatgpt_2024}}
    \label{fig:enter-label}
\end{figure}

According to a recent study, GitHub Copilot, the Microsoft AI assistant, outperformed ChatGPT and Gemini in the same survey. It was also reported to increase developers’ productivity with its detailed code suggestions ~\cite{zhao_github_2023}. OpenAI's ChatGPT scored a median rating of 5.5 on a 6-point scale as a medical assistant, yet ChatGPT's reliability may not always be guaranteed~\cite{huajie_github_2023}. AlphaCode, a code-generating system, participated and ranked above average (in the top 54.3\%) in competitions with more than 5000 participants~\cite{vincent_deepmind_2022}. While these models are precious for obtaining necessary information, avoiding solely relying on them for critical data without fact-checking or consulting a real-life expert is essential (Figure: 2).

\section{ETHICAL ISSUES}

The ethical complications arising from using pre-trained models featured in the language learning models (LLM) like ChatGPT and the GitHub Copilot are undoubtedly prominent~\cite{yan_practical_2024}. The biases in the training data can be inherited by the models with a large dataset due to the algorithm's unconscious bias of replication~\cite{basheer_unleashing_2023}. It is necessary to have safeguards to detect and defeat bias. However, these models without person-specificity cannot leak personal data unless this information has already been collected during training data. However, they may generate answers that sound half-done with sensitive information~\cite{noauthor_safeguarding_nodate}. It, therefore, indicates a random set of generated text with actual information that exists rather than an actual unveiling.

Even though these models are created to give more factual and beneficial information, they can produce convincingly wrong but credible information~\cite{karamolegkou_copyright_2023}. Procedures are still refined in this zone, and a sound test is crucial. Content ownership and copyright may be involved in massive disputes~\cite{kulkarni_google_2023}. Thus, the best example could be written where GitHub copilot creates a core snippet, which might infringe on the original code authors' intellectual property rights~\cite{santosh_large-language-models_2024}. They (Gemini, Alphacode, and Chat GPT) can produce a variety of text forms that are often indistinguishable from original content. Such models need mechanisms to address intellectual property rights and not violate associated forums~\cite{noauthor_gemini_nodate}.

The reality is that such models automate some tasks, which, in turn, might lead people to worry about its effects on the jobs available. In this regard, it should be admitted that the models themselves do not possess the limitless problem-solving capacity of the human brain and can only emulate specific skills. Again, suppose in mental well-being, employing chatbots can give rise to distinct ethical issues. Ethical frameworks should be developed to ensure accountability for these innovations.

\section{FAIRNESS}

All individual models, such as ChatGPT, Gemini, AlphaCode, and GitHub Copilot, share the critical issue of fairness in AI models~\cite{kheya_pursuit_2024}. These approaches offered by business enterprises such as OpenAI and Google offer ideas to increase fairness. However, inconsistencies and errors related to bias detection and generation are still around~\cite{Fang_Che_Mao_Zhang_Zhao_Zhao_2024}. ChatGPT and Gemini have developed a code of conduct to promote impartiality in their answers. However, in some cases, these guidelines may not be followed. The equality of the parties is vital to both AlphaCode and GitHub Copilot since they interact with a vast body of users, and these parties emphasize working to ensure their outputs do not discriminate against any particular group~\cite{dwivedi_opinion_2023}.

The framework evaluates the fairness of LLM and found biased data and social and power structures within the chatbot ChatGPT~\cite{Lippens_2024}. Compared with the earlier GPT-3, there are observed changes through the newer version GPT-4 to a certain extent. However, bias detection in specific contexts remains a problem. An elementary issue when comparing the fairness of Gemini, AlphaCode, and GitHub Copilot metrics is that their fairness standards are less investigatory, which makes a direct comparison hard to do ~\cite{kheya_pursuit_2024}. There are concerns regarding the ethical objectives of AI model development. This aspect might be misidentified due to programmers' biases. There is a perception that, regardless of how serious the problem, the writers do not adequately address these concerns of principle, particularly when developing these models, in contrast to the notion that the staff investments in developing the AI Principles are overly substantial. According to the research findings, fairness concerns have emerged in algorithmic applications such as ChatGPT and Gemini in a variety of fields. The question of whether AlphaCode and human code are of equivalent quality has also been aggressively sought, but no standard benchmark tool is currently available. However, there has been some research that may be useful in reducing gender prejudice induced by LLMs~\cite{Zayed_Mordido_Shabanian_Baldini_Chandar_2024}. We can aspire for future research in this field to eliminate the various sorts of biases created by algorithms, data, selection, labeling, and so on.

\section{LIMITATIONS}

ChatGPT is a language model born due to OpenAI’s efforts, which can handle a broad range of tasks, including translation of texts between languages, language generation via models, and dialogue development among humans. It can create human-like texts and is applicable for completing tasks such as stocking emails, coding, producing content, and others~\cite{cfa_how_2024}. It does not, however, eliminate some drawbacks, such as providing erroneous information, responding inappropriately or incomprehensibly, and having trouble understanding some things. Gemini is a sophisticated artificial intelligence (AI) system developed by Google that is capable of reasoning over a variety of data formats, including words, code used audio, images, and videos~\cite{yan_practical_2024}. It is fast, and every day, packages with the right answers are tapped more closely. However, a number of variables influence the algorithm's performance, which is solely dependent on the caliber and variety of the code it was trained on. The quality and dependability of the resulting code may be impacted if the data used to train it contains bias or gaps~\cite{vincent_deepmind_2022}.GitHub Copilot, a joint creation by GitHub and OpenAI, provides the latter two as an interactive coding assistant in real time~\cite{noauthor_purpose_nodate}. There are different opinions among the users, but it somehow deeply hooks the code environment and performs the task of code completion and suggestion quite professionally~\cite{li_competition-level_2022}. On the one hand, it can help to save time with code writing, but on the other, it can present code suggestions that contain some degree of errors, and improvement by the developer is still needed~\cite{noauthor_transformer_nodate}. Alpha code, designed by DeepMind and named after the first letter of a DNA strand, is a fundamental AI programming model that helps programmers write code that is much more accessible [23]. It has an OpenAI Codex model for the semantic completion of code by replacing functions, entire functions, or giving algorithms in real-time~\cite{dastin_google_2024}. On the other hand, it sometimes fails to perform correctly; for example, it would create a variable and its purpose~\cite{ridnik_code_2024}. Corporations' use of AlphaCode demands substantial computing resources, which the biggest tech companies use.

\section{FUTURE WORK}
At the doorway of a new age in AI, AI models should be at the forefront of the discussion to determine the direction that they will be heading. As Gemini has proven the increase in its implementation, it has been observed to have improved its performance. By comparison, the next-generation model, Gemini 1.5, shows significant cutting-edge changes in all aspects without exception~\cite{wang_gemini_2023}. It has delivered a breakthrough in processing tokens of extended contexts, which are as many as one million. This development will provide unforeseen capabilities and enable developers to bring more valuable models and applications to it. While the future of Google Gemini AI technology remains speculative, the more Google Gemini evolves, the more likely it will be to improve on existing services, providing customers with a more advanced, intuitive, and efficient user. At the same time, GitHub Copilot, focuses on adding support for chat and voice interfaces, handling pull requests, answering questions on docs, and anticipating GPT-4 down the line as the better and more personalized version~\cite{friedman_introducing_2021}. Future software development powered by AI is demonstrated by GitHub Copilot X, the improved version, as the greatest aim of the AI computer. It is meant to revolutionize productivity by eliminating boilerplates and monotonous work, hence simplifying these intricate processes across the developer life cycle. One of the prominent AI tool, ChatGPT has already left its imprint in the computing field, as millions of people use ChatGPT without warning to use it ethically. It is highly likely that as ChatGPT and other language models get more complex and advanced, they will substitute for human workers for multiple tasks. However, they may not necessarily produce 100\% accurate outputs. The code-generating AlphaCode system of DeepMind is capable of creating qualitatively new solutions to complex issues, and past versions, such as AlphaCode 2, have even achieved beating 85\% levels of opponents on average in programming competitions~\cite{Wiggers_2023}. In the future, AlphaCode is anticipated to be a tool competitive programming for programmers~\cite{Castelvecchi_2022}.

The recent achievements in models such as Gemini, ChatGPT, GitHub Copilot, and AlphaCode, which show quick gains and increased application at the dawn of a new era in AI, pave the way for important future developments. On benchmarks like HumanEval and Natural2Code, the Gemini AI models—especially Gemini 1.5 Pro and Gemini-Ultra—have demonstrated remarkable pass@100 rates, demonstrating their increasing capacity to manage challenging tasks with more precision. Similar results have been obtained by ChatGPT's many iterations, such as the GPT-4-Turbo models, which indicate that future iterations will probably continue to improve the system's dependability and efficacy. With its 2024 versions obtaining high test case success rates, GitHub Copilot has also shown efficacy in coding activities, solidifying its place as an increasingly important tool for developers. Even while AlphaCode could still use some refinement in competitive programming, it clearly has the potential to provide creative solutions, and next iterations should build on these strengths. These AI models have the ability to completely change sectors and increase productivity as long as they keep developing. Given these developments, we want to evaluate and further contribute to the continuous development of these state-of-the-art technologies by conducting our own code generation accuracy tests in the future.

\section{CONCLUSION}

Throughout this comprehensive review, we have investigated the architecture, capabilities, and performance of various artificial intelligence models and chatbots, including ChatGPT, Gemini, GitHub Copilot, and AlphaCode, emphasizing their profound impact on language understanding, code generation, and problem solving across a wide range of applications. With ChatGPT producing outstanding results in language processing benchmarks and Gemini demonstrating surprising prowess in tasks like Java code generation, these AI models have completely changed the software development landscape by offering tools that aid engineers in real-time. The fact that GitHub Copilot can offer code recommendations and real-time feedback emphasizes the technologies' revolutionary potential even more. But the review also identifies a number of serious difficulties, such as problems with accuracy, reliability, and morality. Even with these developments, AI systems are not perfect; frequently, they provide results that need human review and revision. The conversation on ethical AI technology use needs to continue, especially as these tools are incorporated more and more into everyday tasks. This study's empirical analysis highlights the present advantages and disadvantages of these AI helpers, emphasizing the need for ongoing development to increase their efficacy and dependability. Metrics such as pass@k and test case pass rates have been used to compare models and provide important insights into how well they perform. ChatGPT and Gemini have shown themselves to be formidable competitors in code generation jobs. The market for AI-powered tools is expected to continue expanding and innovating in the future. In order to ensure that these technologies are refined and suit the changing needs of both sectors and users, it will be essential that we carry out independent testing of code generation accuracy. AI has a bright future in software development and other fields, as long as the problems of ethics, justice, and accuracy are given the attention and concern they require. These technologies will surely become more and more important in determining the direction of programming and other fields as they develop.

\section{Acknowledgement}

The work was partially supported by the National Science Foundation (NSF) Grants CCF-1909963, CNS-2120350, and III-2311598. We thank the NSF for providing us with funding support for this project.

\bibliographystyle{ACM-Reference-Format}
\bibliography{main}


\begin{thebibliography}{74}


\ifx \showCODEN    \undefined \def \showCODEN     #1{\unskip}     \fi
\ifx \showDOI      \undefined \def \showDOI       #1{#1}\fi
\ifx \showISBNx    \undefined \def \showISBNx     #1{\unskip}     \fi
\ifx \showISBNxiii \undefined \def \showISBNxiii  #1{\unskip}     \fi
\ifx \showISSN     \undefined \def \showISSN      #1{\unskip}     \fi
\ifx \showLCCN     \undefined \def \showLCCN      #1{\unskip}     \fi
\ifx \shownote     \undefined \def \shownote      #1{#1}          \fi
\ifx \showarticletitle \undefined \def \showarticletitle #1{#1}   \fi
\ifx \showURL      \undefined \def \showURL       {\relax}        \fi
\providecommand\bibfield[2]{#2}
\providecommand\bibinfo[2]{#2}
\providecommand\natexlab[1]{#1}
\providecommand\showeprint[2][]{arXiv:#2}

\bibitem[\protect\citeauthoryear{??}{noa}{[n.d.]a}]%
        {noauthor_about_nodate}
 \bibinfo{year}{[n.d.]}\natexlab{a}.
\newblock \bibinfo{title}{About {GitHub} Copilot Individual}.
\newblock
\newblock
\urldef\tempurl%
\url{https://docs.github.com/en/copilot/copilot-individual/about-github-copilot-individual}
\showURL{%
\tempurl}


\bibitem[\protect\citeauthoryear{??}{noa}{[n.d.]b}]%
        {noauthor_build_nodate}
 \bibinfo{year}{[n.d.]}\natexlab{b}.
\newblock \bibinfo{title}{Build with the Gemini {API}}.
\newblock
\newblock
\urldef\tempurl%
\url{https://ai.google.dev/}
\showURL{%
\tempurl}


\bibitem[\protect\citeauthoryear{??}{noa}{[n.d.]c}]%
        {noauthor_chatgpt_nodate}
 \bibinfo{year}{[n.d.]}\natexlab{c}.
\newblock \bibinfo{title}{{ChatGPT} vs. Microsoft Copilot vs. Gemini: Which is the best {AI} chatbot?}
\newblock
\newblock
\urldef\tempurl%
\url{https://www.zdnet.com/article/chatgpt-vs-microsoft-copilot-vs-gemini-which-is-the-best-ai-chatbot/}
\showURL{%
\tempurl}


\bibitem[\protect\citeauthoryear{??}{noa}{[n.d.]d}]%
        {noauthor_chord_nodate}
 \bibinfo{year}{[n.d.]}\natexlab{d}.
\newblock \bibinfo{title}{Chord}.
\newblock
\newblock
\urldef\tempurl%
\url{https://chord.pub/article/39449/how-to-use-alphacode}
\showURL{%
\tempurl}


\bibitem[\protect\citeauthoryear{??}{noa}{[n.d.]e}]%
        {noauthor_code_2023}
 \bibinfo{year}{[n.d.]}\natexlab{e}.
\newblock \bibinfo{title}{Code and debug with Bard}.
\newblock
\newblock
\urldef\tempurl%
\url{https://blog.google/technology/ai/code-with-bard/}
\showURL{%
\tempurl}


\bibitem[\protect\citeauthoryear{??}{noa}{[n.d.]f}]%
        {noauthor_competitive_2022}
 \bibinfo{year}{[n.d.]}\natexlab{f}.
\newblock \bibinfo{title}{Competitive programming with {AlphaCode}}.
\newblock
\newblock
\urldef\tempurl%
\url{https://deepmind.google/discover/blog/competitive-programming-with-alphacode/}
\showURL{%
\tempurl}


\bibitem[\protect\citeauthoryear{??}{noa}{[n.d.]g}]%
        {noauthor_gemini_nodate}
 \bibinfo{year}{[n.d.]}\natexlab{g}.
\newblock \bibinfo{title}{Gemini - Google {DeepMind}}.
\newblock
\newblock
\urldef\tempurl%
\url{https://deepmind.google/technologies/gemini/}
\showURL{%
\tempurl}


\bibitem[\protect\citeauthoryear{??}{noa}{[n.d.]h}]%
        {noauthor_gemini_2016}
 \bibinfo{year}{[n.d.]}\natexlab{h}.
\newblock \bibinfo{title}{Gemini: All About This Zodiac Sign's Personality Traits, Compatibility and More}.
\newblock
\newblock
\urldef\tempurl%
\url{https://astrostyle.com/astrology/zodiac-signs/gemini/}
\showURL{%
\tempurl}


\bibitem[\protect\citeauthoryear{??}{noa}{[n.d.]i}]%
        {noauthor_getting_nodate}
 \bibinfo{year}{[n.d.]}\natexlab{i}.
\newblock \bibinfo{title}{Getting started with {GitHub} Copilot}.
\newblock
\newblock
\urldef\tempurl%
\url{https://docs.github.com/en/copilot/using-github-copilot/getting-started-with-github-copilot}
\showURL{%
\tempurl}


\bibitem[\protect\citeauthoryear{??}{noa}{[n.d.]j}]%
        {noauthor_github_nodate}
 \bibinfo{year}{[n.d.]}\natexlab{j}.
\newblock \showarticletitle{{GitHub} Copilot vs. {ChatGPT}: Which is Better for Coding in 2024?}
\newblock  (\bibinfo{year}{[n.\,d.]}).
\newblock


\bibitem[\protect\citeauthoryear{??}{noa}{[n.d.]k}]%
        {noauthor_gpt-35_2023}
 \bibinfo{year}{[n.d.]}\natexlab{k}.
\newblock \bibinfo{title}{{GPT}-3.5 model architecture}.
\newblock
\newblock
\urldef\tempurl%
\url{https://iq.opengenus.org/gpt-3-5-model/}
\showURL{%
\tempurl}


\bibitem[\protect\citeauthoryear{??}{noa}{[n.d.]l}]%
        {noauthor_models_nodate}
 \bibinfo{year}{[n.d.]}\natexlab{l}.
\newblock \bibinfo{title}{Models comparison: {OpenAI} documentation}.
\newblock
\newblock
\urldef\tempurl%
\url{https://platform.openai.com/docs/models/overview}
\showURL{%
\tempurl}


\bibitem[\protect\citeauthoryear{??}{noa}{[n.d.]m}]%
        {noauthor_plm_2023}
 \bibinfo{year}{[n.d.]}\natexlab{m}.
\newblock \bibinfo{title}{{PLM}, {ChatGPT}, and Large Language Model Thoughts}.
\newblock
\newblock
\urldef\tempurl%
\url{https://beyondplm.com/2023/01/28/plm-chatgpt-and-large-language-model-thoughts/}
\showURL{%
\tempurl}


\bibitem[\protect\citeauthoryear{??}{noa}{[n.d.]n}]%
        {noauthor_purpose_nodate}
 \bibinfo{year}{[n.d.]}\natexlab{n}.
\newblock \bibinfo{title}{The purpose, benefits, and downsides of {GitHub} Copilot {\textbar} Proxify.io}.
\newblock
\newblock
\urldef\tempurl%
\url{https://proxify.io/articles/what-is-github-copilot}
\showURL{%
\tempurl}


\bibitem[\protect\citeauthoryear{??}{noa}{[n.d.]o}]%
        {noauthor_quickstart_nodate}
 \bibinfo{year}{[n.d.]}\natexlab{o}.
\newblock \bibinfo{title}{Quickstart for {GitHub} Copilot}.
\newblock
\newblock
\urldef\tempurl%
\url{https://docs.github.com/en/copilot/quickstart}
\showURL{%
\tempurl}


\bibitem[\protect\citeauthoryear{??}{noa}{[n.d.]p}]%
        {noauthor_safeguarding_nodate}
 \bibinfo{year}{[n.d.]}\natexlab{p}.
\newblock \bibinfo{title}{Safeguarding Data Integrity and Privacy in the Age of {LLMs} {\textbar} Sentra Blog}.
\newblock
\newblock
\urldef\tempurl%
\url{https://www.sentra.io/blog/safeguarding-data-integrity-and-privacy-in-the-age-of-ai-powered-large-language-models-llms}
\showURL{%
\tempurl}


\bibitem[\protect\citeauthoryear{??}{noa}{[n.d.]q}]%
        {noauthor_transformer_nodate}
 \bibinfo{year}{[n.d.]}\natexlab{q}.
\newblock \bibinfo{title}{The transformer architecture {\textbar} Python}.
\newblock
\newblock
\urldef\tempurl%
\url{https://campus.datacamp.com/courses/introduction-to-llms-in-python/the-large-language-models-llms-landscape?ex=7}
\showURL{%
\tempurl}


\bibitem[\protect\citeauthoryear{??}{noa}{[n.d.]r}]%
        {noauthor_understanding_2024}
 \bibinfo{year}{[n.d.]}\natexlab{r}.
\newblock \bibinfo{title}{Understanding Transformers \& the Architecture of {LLMs}}.
\newblock
\newblock
\urldef\tempurl%
\url{https://www.mlq.ai/llm-transformer-architecture/}
\showURL{%
\tempurl}


\bibitem[\protect\citeauthoryear{??}{noa}{[n.d.]s}]%
        {noauthor_what_2024}
 \bibinfo{year}{[n.d.]}\natexlab{s}.
\newblock \bibinfo{title}{What Are Large Language Models ({LLMs})? {\textbar} {IBM}}.
\newblock
\newblock
\urldef\tempurl%
\url{https://www.ibm.com/topics/large-language-models}
\showURL{%
\tempurl}


\bibitem[\protect\citeauthoryear{??}{noa}{[n.d.]t}]%
        {noauthor_what_nodate}
 \bibinfo{year}{[n.d.]}\natexlab{t}.
\newblock \bibinfo{title}{What is {ChatGPT} and why does it matter? Here's what you need to know}.
\newblock
\newblock
\urldef\tempurl%
\url{https://www.zdnet.com/article/what-is-chatgpt-and-why-does-it-matter-heres-everything-you-need-to-know/}
\showURL{%
\tempurl}


\bibitem[\protect\citeauthoryear{??}{Our}{2024}]%
        {Our_next_generation_model_2024}
 \bibinfo{year}{2024}\natexlab{}.
\newblock
\newblock
\urldef\tempurl%
\url{https://blog.google/technology/ai/google-gemini-next-generation-model-february-2024/}
\showURL{%
\tempurl}


\bibitem[\protect\citeauthoryear{Akter et~al\mbox{.}}{Akter et~al\mbox{.}}{2023}]%
        {Akter_Yu_Muhamed_Ou_Bäuerle_Cabrera_Dholakia_Xiong_Neubig_2023}
\bibfield{author}{\bibinfo{person}{Syeda~Nahida Akter} {et~al\mbox{.}}} \bibinfo{year}{2023}\natexlab{}.
\newblock \showarticletitle{An In-depth Look at Gemini’s Language Abilities}.
\newblock  (\bibinfo{year}{2023}).
\newblock
\urldef\tempurl%
\url{https://doi.org/10.48550/ARXIV.2312.11444}
\showDOI{\tempurl}


\bibitem[\protect\citeauthoryear{Basheer}{Basheer}{[n.d.]}]%
        {basheer_unleashing_2023}
\bibfield{author}{\bibinfo{person}{K.~C.~Sabreena Basheer}.} \bibinfo{year}{[n.d.]}\natexlab{}.
\newblock \bibinfo{title}{Unleashing the Power of {DeepMind}'s {AlphaCode}: Revolutionizing Code Writing}.
\newblock
\newblock
\urldef\tempurl%
\url{https://www.analyticsvidhya.com/blog/2023/12/unleashing-the-power-of-deepminds-alphacode-revolutionizing-code-writing/}
\showURL{%
\tempurl}


\bibitem[\protect\citeauthoryear{Bengio, Ducharme, and Vincent}{Bengio et~al\mbox{.}}{2000}]%
        {NIPS2000_728f206c}
\bibfield{author}{\bibinfo{person}{Yoshua Bengio}, \bibinfo{person}{R\'{e}jean Ducharme}, {and} \bibinfo{person}{Pascal Vincent}.} \bibinfo{year}{2000}\natexlab{}.
\newblock \showarticletitle{A Neural Probabilistic Language Model}. In \bibinfo{booktitle}{\emph{Advances in Neural Information Processing Systems}}, \bibfield{editor}{\bibinfo{person}{T.~Leen}, \bibinfo{person}{T.~Dietterich}, {and} \bibinfo{person}{V.~Tresp}} (Eds.), Vol.~\bibinfo{volume}{13}. \bibinfo{publisher}{MIT Press}.
\newblock
\urldef\tempurl%
\url{https://proceedings.neurips.cc/paper_files/paper/2000/file/728f206c2a01bf572b5940d7d9a8fa4c-Paper.pdf}
\showURL{%
\tempurl}


\bibitem[\protect\citeauthoryear{Castelvecchi}{Castelvecchi}{2022}]%
        {Castelvecchi_2022}
\bibfield{author}{\bibinfo{person}{Davide Castelvecchi}.} \bibinfo{year}{2022}\natexlab{}.
\newblock \showarticletitle{Are ChatGPT and AlphaCode going to replace programmers?}
\newblock \bibinfo{journal}{\emph{Nature}} (\bibinfo{date}{Dec.} \bibinfo{year}{2022}).
\newblock
\urldef\tempurl%
\url{https://doi.org/10.1038/d41586-022-04383-z}
\showDOI{\tempurl}


\bibitem[\protect\citeauthoryear{{CFA}}{{CFA}}{[n.d.]}]%
        {cfa_how_2024}
\bibfield{author}{\bibinfo{person}{Sam~{McKay} {CFA}}.} \bibinfo{year}{[n.d.]}\natexlab{}.
\newblock \bibinfo{title}{How to Use Chat {GPT}: A Simple Guide for Beginners {\textbar} Master Data Skills + {AI}}.
\newblock
\newblock
\urldef\tempurl%
\url{https://blog.enterprisedna.co/how-to-use-chat-gpt/}
\showURL{%
\tempurl}


\bibitem[\protect\citeauthoryear{Chandra, Tünnermann, Löfstedt, and Gratz}{Chandra et~al\mbox{.}}{[n.d.]}]%
        {chandra_transformer-based_2023}
\bibfield{author}{\bibinfo{person}{Abel Chandra}, \bibinfo{person}{Laura Tünnermann}, \bibinfo{person}{Tommy Löfstedt}, {and} \bibinfo{person}{Regina Gratz}.} \bibinfo{year}{[n.d.]}\natexlab{}.
\newblock \showarticletitle{Transformer-based deep learning for predicting protein properties in the life sciences}.
\newblock   \bibinfo{volume}{12} (\bibinfo{year}{[n.\,d.]}), \bibinfo{pages}{e82819}.
\newblock
\showISSN{2050-084X}
\urldef\tempurl%
\url{https://doi.org/10.7554/eLife.82819}
\showDOI{\tempurl}


\bibitem[\protect\citeauthoryear{Dastin and Dastin}{Dastin and Dastin}{[n.d.]a}]%
        {dastin_google_2024}
\bibfield{author}{\bibinfo{person}{Jeffrey Dastin} {and} \bibinfo{person}{Jeffrey Dastin}.} \bibinfo{year}{[n.d.]}\natexlab{a}.
\newblock \showarticletitle{Google rebrands Bard chatbot as Gemini, rolls out paid subscription}.
\newblock  (\bibinfo{year}{[n.\,d.]}).
\newblock
\urldef\tempurl%
\url{https://www.reuters.com/technology/google-rebrands-bard-chatbot-gemini-rolls-out-paid-subscription-2024-02-08/}
\showURL{%
\tempurl}


\bibitem[\protect\citeauthoryear{Dastin and Dastin}{Dastin and Dastin}{[n.d.]b}]%
        {dastin_google_2023}
\bibfield{author}{\bibinfo{person}{Jeffrey Dastin} {and} \bibinfo{person}{Jeffrey Dastin}.} \bibinfo{year}{[n.d.]}\natexlab{b}.
\newblock \showarticletitle{Google unveils {ChatGPT} rival Bard, {AI} search plans in battle with Microsoft}.
\newblock  (\bibinfo{year}{[n.\,d.]}).
\newblock
\urldef\tempurl%
\url{https://www.reuters.com/technology/google-opens-bard-chatbot-test-users-plans-more-ai-search-2023-02-06/}
\showURL{%
\tempurl}


\bibitem[\protect\citeauthoryear{Devlin, Chang, Lee, and Toutanova}{Devlin et~al\mbox{.}}{2019}]%
        {devlin-etal-2019-bert}
\bibfield{author}{\bibinfo{person}{Jacob Devlin}, \bibinfo{person}{Ming-Wei Chang}, \bibinfo{person}{Kenton Lee}, {and} \bibinfo{person}{Kristina Toutanova}.} \bibinfo{year}{2019}\natexlab{}.
\newblock \showarticletitle{{BERT}: Pre-training of Deep Bidirectional Transformers for Language Understanding}. In \bibinfo{booktitle}{\emph{Proceedings of the 2019 Conference of the North {A}merican Chapter of the Association for Computational Linguistics: Human Language Technologies, Volume 1 (Long and Short Papers)}}, \bibfield{editor}{\bibinfo{person}{Jill Burstein}, \bibinfo{person}{Christy Doran}, {and} \bibinfo{person}{Thamar Solorio}} (Eds.). \bibinfo{publisher}{Association for Computational Linguistics}, \bibinfo{address}{Minneapolis, Minnesota}, \bibinfo{pages}{4171--4186}.
\newblock
\urldef\tempurl%
\url{https://doi.org/10.18653/v1/N19-1423}
\showDOI{\tempurl}


\bibitem[\protect\citeauthoryear{Dibia}{Dibia}{[n.d.]}]%
        {dibia_alphacode_2022}
\bibfield{author}{\bibinfo{person}{Victor Dibia}.} \bibinfo{year}{[n.d.]}\natexlab{}.
\newblock \bibinfo{title}{{AlphaCode}: Competition-Level Code Generation with Transformer Based Architectures {\textbar} Paper Review}.
\newblock
\newblock
\urldef\tempurl%
\url{https://victordibia.com}
\showURL{%
\tempurl}


\bibitem[\protect\citeauthoryear{Dohmke}{Dohmke}{[n.d.]}]%
        {dohmke_github_2022}
\bibfield{author}{\bibinfo{person}{Thomas Dohmke}.} \bibinfo{year}{[n.d.]}\natexlab{}.
\newblock \bibinfo{title}{{GitHub} Copilot is generally available to all developers}.
\newblock
\newblock
\urldef\tempurl%
\url{https://github.blog/2022-06-21-github-copilot-is-generally-available-to-all-developers/}
\showURL{%
\tempurl}


\bibitem[\protect\citeauthoryear{Dwivedi et~al\mbox{.}}{Dwivedi et~al\mbox{.}}{[n.d.]}]%
        {dwivedi_opinion_2023}
\bibfield{author}{\bibinfo{person}{Yogesh~K. Dwivedi} {et~al\mbox{.}}} \bibinfo{year}{[n.d.]}\natexlab{}.
\newblock \showarticletitle{Opinion Paper: “So what if {ChatGPT} wrote it?” Multidisciplinary perspectives on opportunities, challenges and implications of generative conversational {AI} for research, practice and policy}.
\newblock   \bibinfo{volume}{71} (\bibinfo{year}{[n.\,d.]}), \bibinfo{pages}{102642}.
\newblock
\showISSN{0268-4012}
\urldef\tempurl%
\url{https://doi.org/10.1016/j.ijinfomgt.2023.102642}
\showDOI{\tempurl}


\bibitem[\protect\citeauthoryear{Fang, Che, Mao, Zhang, Zhao, and Zhao}{Fang et~al\mbox{.}}{2024}]%
        {Fang_Che_Mao_Zhang_Zhao_Zhao_2024}
\bibfield{author}{\bibinfo{person}{Xiao Fang}, \bibinfo{person}{Shangkun Che}, \bibinfo{person}{Minjia Mao}, \bibinfo{person}{Hongzhe Zhang}, \bibinfo{person}{Ming Zhao}, {and} \bibinfo{person}{Xiaohang Zhao}.} \bibinfo{year}{2024}\natexlab{}.
\newblock \showarticletitle{Bias of AI-generated content: an examination of news produced by large language models}.
\newblock \bibinfo{journal}{\emph{Scientific Reports}} \bibinfo{volume}{14}, \bibinfo{number}{1} (\bibinfo{date}{March} \bibinfo{year}{2024}), \bibinfo{pages}{5224}.
\newblock
\urldef\tempurl%
\url{https://doi.org/10.1038/s41598-024-55686-2}
\showDOI{\tempurl}


\bibitem[\protect\citeauthoryear{Friedman}{Friedman}{[n.d.]}]%
        {friedman_introducing_2021}
\bibfield{author}{\bibinfo{person}{Nat Friedman}.} \bibinfo{year}{[n.d.]}\natexlab{}.
\newblock \bibinfo{title}{Introducing {GitHub} Copilot: your {AI} pair programmer}.
\newblock
\newblock
\urldef\tempurl%
\url{https://github.blog/2021-06-29-introducing-github-copilot-ai-pair-programmer/}
\showURL{%
\tempurl}


\bibitem[\protect\citeauthoryear{Gates}{Gates}{[n.d.]}]%
        {gates_age_nodate}
\bibfield{author}{\bibinfo{person}{Bill Gates}.} \bibinfo{year}{[n.d.]}\natexlab{}.
\newblock \bibinfo{title}{The Age of {AI} has begun}.
\newblock
\newblock
\urldef\tempurl%
\url{https://www.gatesnotes.com/The-Age-of-AI-Has-Begun}
\showURL{%
\tempurl}


\bibitem[\protect\citeauthoryear{Gershgorn}{Gershgorn}{[n.d.]}]%
        {gershgorn_github_2021}
\bibfield{author}{\bibinfo{person}{Dave Gershgorn}.} \bibinfo{year}{[n.d.]}\natexlab{}.
\newblock \bibinfo{title}{{GitHub} and {OpenAI} launch a new {AI} tool that generates its own code}.
\newblock
\newblock
\urldef\tempurl%
\url{https://www.theverge.com/2021/6/29/22555777/github-openai-ai-tool-autocomplete-code}
\showURL{%
\tempurl}


\bibitem[\protect\citeauthoryear{Hines}{Hines}{[n.d.]}]%
        {hines_history_2023}
\bibfield{author}{\bibinfo{person}{Kristi Hines}.} \bibinfo{year}{[n.d.]}\natexlab{}.
\newblock \bibinfo{title}{History Of {ChatGPT}: A Timeline Of The Meteoric Rise Of Generative {AI} Chatbots}.
\newblock
\newblock
\urldef\tempurl%
\url{https://www.searchenginejournal.com/history-of-chatgpt-timeline/488370/}
\showURL{%
\tempurl}


\bibitem[\protect\citeauthoryear{Huajie}{Huajie}{[n.d.]}]%
        {huajie_github_2023}
\bibfield{author}{\bibinfo{person}{Xu Huajie}.} \bibinfo{year}{[n.d.]}\natexlab{}.
\newblock \showarticletitle{Github Copilot - A Groundbreaking Code Autocomplete Tool}.
\newblock  (\bibinfo{year}{[n.\,d.]}).
\newblock
\urldef\tempurl%
\url{https://doi.org/10.13140/RG.2.2.29962.24002}
\showDOI{\tempurl}


\bibitem[\protect\citeauthoryear{Isci}{Isci}{[n.d.]}]%
        {isci_comprehensive_2023}
\bibfield{author}{\bibinfo{person}{Senol Isci}.} \bibinfo{year}{[n.d.]}\natexlab{}.
\newblock \bibinfo{title}{Comprehensive Guide on Evaluation of Response Generation and Retrieval in {LLMs}}.
\newblock
\newblock
\urldef\tempurl%
\url{https://medium.com/@senol.isci/comprehensive-guide-on-evaluation-of-response-generation-and-retrieval-with-llms-0cbc2adb3ae6}
\showURL{%
\tempurl}


\bibitem[\protect\citeauthoryear{Jelinek}{Jelinek}{[n.d.]}]%
        {jelinek_statistical_1997}
\bibfield{author}{\bibinfo{person}{Frederick Jelinek}.} \bibinfo{year}{[n.d.]}\natexlab{}.
\newblock \bibinfo{booktitle}{\emph{Statistical methods for speech recognition}}.
\newblock \bibinfo{publisher}{{MIT} Press}.
\newblock
\showISBNx{9780262100663}


\bibitem[\protect\citeauthoryear{Kaftan, Hussain, and Naser}{Kaftan et~al\mbox{.}}{[n.d.]}]%
        {kaftan_response_2024}
\bibfield{author}{\bibinfo{person}{Ahmed~Naseer Kaftan}, \bibinfo{person}{Majid~Kadhum Hussain}, {and} \bibinfo{person}{Farah~Hasson Naser}.} \bibinfo{year}{[n.d.]}\natexlab{}.
\newblock \showarticletitle{Response accuracy of {ChatGPT} 3.5 Copilot and Gemini in interpreting biochemical laboratory data a pilot study}.
\newblock  \bibinfo{volume}{14}, \bibinfo{number}{1} (\bibinfo{year}{[n.\,d.]}), \bibinfo{pages}{8233}.
\newblock
\showISSN{2045-2322}
\urldef\tempurl%
\url{https://doi.org/10.1038/s41598-024-58964-1}
\showDOI{\tempurl}


\bibitem[\protect\citeauthoryear{Karamolegkou, Li, Zhou, and Søgaard}{Karamolegkou et~al\mbox{.}}{[n.d.]}]%
        {karamolegkou_copyright_2023}
\bibfield{author}{\bibinfo{person}{Antonia Karamolegkou}, \bibinfo{person}{Jiaang Li}, \bibinfo{person}{Li Zhou}, {and} \bibinfo{person}{Anders Søgaard}.} \bibinfo{year}{[n.d.]}\natexlab{}.
\newblock \bibinfo{title}{Copyright Violations and Large Language Models}.
\newblock
\newblock
\urldef\tempurl%
\url{https://doi.org/10.48550/arXiv.2310.13771}
\showDOI{\tempurl}
\showeprint[arxiv]{2310.13771 [cs]}


\bibitem[\protect\citeauthoryear{Kheya, Bouadjenek, and Aryal}{Kheya et~al\mbox{.}}{[n.d.]}]%
        {kheya_pursuit_2024}
\bibfield{author}{\bibinfo{person}{Tahsin~Alamgir Kheya}, \bibinfo{person}{Mohamed~Reda Bouadjenek}, {and} \bibinfo{person}{Sunil Aryal}.} \bibinfo{year}{[n.d.]}\natexlab{}.
\newblock \bibinfo{title}{The Pursuit of Fairness in Artificial Intelligence Models: A Survey}.
\newblock
\newblock
\showeprint[arxiv]{2403.17333 [cs]}
\urldef\tempurl%
\url{http://arxiv.org/abs/2403.17333}
\showURL{%
\tempurl}


\bibitem[\protect\citeauthoryear{Kombrink, Mikolov, Karafiát, and Burget}{Kombrink et~al\mbox{.}}{2011}]%
        {kombrink11_interspeech}
\bibfield{author}{\bibinfo{person}{Stefan Kombrink}, \bibinfo{person}{Tomáš Mikolov}, \bibinfo{person}{Martin Karafiát}, {and} \bibinfo{person}{Lukáš Burget}.} \bibinfo{year}{2011}\natexlab{}.
\newblock \showarticletitle{{Recurrent neural network based language modeling in meeting recognition}}. In \bibinfo{booktitle}{\emph{Proc. Interspeech 2011}}. \bibinfo{pages}{2877--2880}.
\newblock
\showISSN{2308-457X}
\urldef\tempurl%
\url{https://doi.org/10.21437/Interspeech.2011-720}
\showDOI{\tempurl}


\bibitem[\protect\citeauthoryear{Kulkarni, Shivananda, Kulkarni, and Gudivada}{Kulkarni et~al\mbox{.}}{[n.d.]}]%
        {kulkarni_google_2023}
\bibfield{author}{\bibinfo{person}{Akshay Kulkarni}, \bibinfo{person}{Adarsha Shivananda}, \bibinfo{person}{Anoosh Kulkarni}, {and} \bibinfo{person}{Dilip Gudivada}.} \bibinfo{year}{[n.d.]}\natexlab{}.
\newblock \bibinfo{booktitle}{\emph{Google Bard and Beyond}}.
\newblock \bibinfo{publisher}{Apress}, \bibinfo{pages}{79--99}.
\newblock
\showISBNx{9781484299937 9781484299944}
\urldef\tempurl%
\url{https://doi.org/10.1007/978-1-4842-9994-4_5}
\showDOI{\tempurl}


\bibitem[\protect\citeauthoryear{Kumar, Srivastava, Dwivedi, Budhiraja, Ghosh, Goyal, and Arora}{Kumar et~al\mbox{.}}{[n.d.]}]%
        {santosh_large-language-models_2024}
\bibfield{author}{\bibinfo{person}{Vimal Kumar}, \bibinfo{person}{Priyam Srivastava}, \bibinfo{person}{Ashay Dwivedi}, \bibinfo{person}{Ishan Budhiraja}, \bibinfo{person}{Debjani Ghosh}, \bibinfo{person}{Vikas Goyal}, {and} \bibinfo{person}{Ruchika Arora}.} \bibinfo{year}{[n.d.]}\natexlab{}.
\newblock \showarticletitle{Large-Language-Models ({LLM})-Based {AI} Chatbots: Architecture, In-Depth Analysis and Their Performance Evaluation}.
\newblock In \bibinfo{booktitle}{\emph{Recent Trends in Image Processing and Pattern Recognition}}, \bibfield{editor}{\bibinfo{person}{Kc~Santosh}, \bibinfo{person}{Aaisha Makkar}, \bibinfo{person}{Myra Conway}, \bibinfo{person}{Ashutosh~K. Singh}, \bibinfo{person}{Antoine Vacavant}, \bibinfo{person}{Anas Abou El~Kalam}, \bibinfo{person}{Mohamed-Rafik Bouguelia}, {and} \bibinfo{person}{Ravindra Hegadi}} (Eds.). Vol.~\bibinfo{volume}{2027}. \bibinfo{publisher}{Springer Nature Switzerland}, \bibinfo{pages}{237--249}.
\newblock
\showISBNx{9783031530845 9783031530852}
\urldef\tempurl%
\url{https://doi.org/10.1007/978-3-031-53085-2_20}
\showDOI{\tempurl}


\bibitem[\protect\citeauthoryear{Li et~al\mbox{.}}{Li et~al\mbox{.}}{2022}]%
        {Li_Choi_Chung_Kushman}
\bibfield{author}{\bibinfo{person}{Yujia Li} {et~al\mbox{.}}} \bibinfo{year}{2022}\natexlab{}.
\newblock \showarticletitle{Competition-level code generation with AlphaCode}.
\newblock \bibinfo{journal}{\emph{Science}} \bibinfo{volume}{378}, \bibinfo{number}{6624} (\bibinfo{date}{Dec.} \bibinfo{year}{2022}), \bibinfo{pages}{1092–1097}.
\newblock
\urldef\tempurl%
\url{https://doi.org/10.1126/science.abq1158}
\showDOI{\tempurl}


\bibitem[\protect\citeauthoryear{Li, Choi, Chung, Kushman, Schrittwieser, Leblond, Eccles, Keeling, Gimeno, Dal~Lago, Hubert, Choy, De~Masson~d’Autume, Babuschkin, Chen, Huang, Welbl, Gowal, Cherepanov, Molloy, Mankowitz, Sutherland~Robson, Kohli, De~Freitas, Kavukcuoglu, and Vinyals}{Li et~al\mbox{.}}{[n.d.]}]%
        {li_competition-level_2022}
\bibfield{author}{\bibinfo{person}{Yujia Li}, \bibinfo{person}{David Choi}, \bibinfo{person}{Junyoung Chung}, \bibinfo{person}{Nate Kushman}, \bibinfo{person}{Julian Schrittwieser}, \bibinfo{person}{Rémi Leblond}, \bibinfo{person}{Tom Eccles}, \bibinfo{person}{James Keeling}, \bibinfo{person}{Felix Gimeno}, \bibinfo{person}{Agustin Dal~Lago}, \bibinfo{person}{Thomas Hubert}, \bibinfo{person}{Peter Choy}, \bibinfo{person}{Cyprien De~Masson~d’Autume}, \bibinfo{person}{Igor Babuschkin}, \bibinfo{person}{Xinyun Chen}, \bibinfo{person}{Po-Sen Huang}, \bibinfo{person}{Johannes Welbl}, \bibinfo{person}{Sven Gowal}, \bibinfo{person}{Alexey Cherepanov}, \bibinfo{person}{James Molloy}, \bibinfo{person}{Daniel~J. Mankowitz}, \bibinfo{person}{Esme Sutherland~Robson}, \bibinfo{person}{Pushmeet Kohli}, \bibinfo{person}{Nando De~Freitas}, \bibinfo{person}{Koray Kavukcuoglu}, {and} \bibinfo{person}{Oriol Vinyals}.} \bibinfo{year}{[n.d.]}\natexlab{}.
\newblock \showarticletitle{Competition-level code generation with {AlphaCode}}.
\newblock  \bibinfo{volume}{378}, \bibinfo{number}{6624} (\bibinfo{year}{[n.\,d.]}), \bibinfo{pages}{1092--1097}.
\newblock
\showISSN{0036-8075, 1095-9203}
\urldef\tempurl%
\url{https://doi.org/10.1126/science.abq1158}
\showDOI{\tempurl}


\bibitem[\protect\citeauthoryear{Lippens}{Lippens}{2024}]%
        {Lippens_2024}
\bibfield{author}{\bibinfo{person}{Louis Lippens}.} \bibinfo{year}{2024}\natexlab{}.
\newblock \showarticletitle{Computer says ‘no’: Exploring systemic bias in ChatGPT using an audit approach}.
\newblock \bibinfo{journal}{\emph{Computers in Human Behavior: Artificial Humans}} \bibinfo{volume}{2}, \bibinfo{number}{1} (\bibinfo{date}{Jan.} \bibinfo{year}{2024}), \bibinfo{pages}{100054}.
\newblock
\urldef\tempurl%
\url{https://doi.org/10.1016/j.chbah.2024.100054}
\showDOI{\tempurl}


\bibitem[\protect\citeauthoryear{Liu et~al\mbox{.}}{Liu et~al\mbox{.}}{[n.d.]}]%
        {liu_refining_2024}
\bibfield{author}{\bibinfo{person}{Yue Liu} {et~al\mbox{.}}} \bibinfo{year}{[n.d.]}\natexlab{}.
\newblock \showarticletitle{Refining {ChatGPT}-Generated Code: Characterizing and Mitigating Code Quality Issues}.
\newblock  \bibinfo{volume}{33}, \bibinfo{number}{5} (\bibinfo{year}{[n.\,d.]}), \bibinfo{pages}{1--26}.
\newblock
\showISSN{1049-331X, 1557-7392}
\urldef\tempurl%
\url{https://doi.org/10.1145/3643674}
\showDOI{\tempurl}


\bibitem[\protect\citeauthoryear{Ma, Liu, Lin, Wang, Hu, Liu, Zhang, Nie, Li, and Liu}{Ma et~al\mbox{.}}{[n.d.]}]%
        {ma_lms_2023}
\bibfield{author}{\bibinfo{person}{Wei Ma}, \bibinfo{person}{Shangqing Liu}, \bibinfo{person}{Zhihao Lin}, \bibinfo{person}{Wenhan Wang}, \bibinfo{person}{Qiang Hu}, \bibinfo{person}{Ye Liu}, \bibinfo{person}{Cen Zhang}, \bibinfo{person}{Liming Nie}, \bibinfo{person}{Li Li}, {and} \bibinfo{person}{Yang Liu}.} \bibinfo{year}{[n.d.]}\natexlab{}.
\newblock \showarticletitle{{LMs}: Understanding Code Syntax and Semantics for Code Analysis}.
\newblock  (\bibinfo{year}{[n.\,d.]}).
\newblock
\urldef\tempurl%
\url{https://doi.org/10.48550/ARXIV.2305.12138}
\showDOI{\tempurl}


\bibitem[\protect\citeauthoryear{Milmo}{Milmo}{[n.d.]}]%
        {milmo_chatgpt_2023}
\bibfield{author}{\bibinfo{person}{Dan Milmo}.} \bibinfo{year}{[n.d.]}\natexlab{}.
\newblock \showarticletitle{{ChatGPT} reaches 100 million users two months after launch}.
\newblock  (\bibinfo{year}{[n.\,d.]}).
\newblock
\showISSN{0261-3077}
\urldef\tempurl%
\url{https://www.theguardian.com/technology/2023/feb/02/chatgpt-100-million-users-open-ai-fastest-growing-app}
\showURL{%
\tempurl}


\bibitem[\protect\citeauthoryear{Mollick}{Mollick}{[n.d.]}]%
        {mollick_chatgpt_2022}
\bibfield{author}{\bibinfo{person}{Ethan Mollick}.} \bibinfo{year}{[n.d.]}\natexlab{}.
\newblock \showarticletitle{{ChatGPT} Is a Tipping Point for {AI}}.
\newblock  (\bibinfo{year}{[n.\,d.]}).
\newblock
\showISSN{0017-8012}
\urldef\tempurl%
\url{https://hbr.org/2022/12/chatgpt-is-a-tipping-point-for-ai}
\showURL{%
\tempurl}


\bibitem[\protect\citeauthoryear{Nguyen and Nadi}{Nguyen and Nadi}{2022}]%
        {Nguyen_Nadi_2022}
\bibfield{author}{\bibinfo{person}{Nhan Nguyen} {and} \bibinfo{person}{Sarah Nadi}.} \bibinfo{year}{2022}\natexlab{}.
\newblock \showarticletitle{An empirical evaluation of GitHub copilot’s code suggestions}. In \bibinfo{booktitle}{\emph{Proceedings of the 19th International Conference on Mining Software Repositories}}. \bibinfo{publisher}{ACM}, \bibinfo{address}{Pittsburgh Pennsylvania}, \bibinfo{pages}{1–5}.
\newblock
\showISBNx{9781450393034}
\urldef\tempurl%
\url{https://doi.org/10.1145/3524842.3528470}
\showDOI{\tempurl}


\bibitem[\protect\citeauthoryear{{OpenAI}}{{OpenAI}}{[n.d.]}]%
        {openai_chatgpt_2024}
\bibfield{author}{\bibinfo{person}{{OpenAI}}.} \bibinfo{year}{[n.d.]}\natexlab{}.
\newblock \bibinfo{title}{Code generated by ChatGPT}.
\newblock
\newblock
\urldef\tempurl%
\url{https://chat.openai.com}
\showURL{%
\tempurl}
\newblock
\shownote{Generated by ChatGPT.}


\bibitem[\protect\citeauthoryear{OpenAI, Achiam, et~al\mbox{.}}{OpenAI et~al\mbox{.}}{2023}]%
        {OpenAI_Achiam_Adler_Agarwal_Ahmad_Akkaya_Aleman_Almeida_Altenschmidt_Altman_et_al_2023}
\bibfield{author}{\bibinfo{person}{OpenAI}, \bibinfo{person}{Josh Achiam}, {et~al\mbox{.}}} \bibinfo{year}{2023}\natexlab{}.
\newblock \showarticletitle{GPT-4 Technical Report}.
\newblock  (\bibinfo{year}{2023}).
\newblock
\urldef\tempurl%
\url{https://doi.org/10.48550/ARXIV.2303.08774}
\showDOI{\tempurl}


\bibitem[\protect\citeauthoryear{Paul, Zhu, and Bayley}{Paul et~al\mbox{.}}{2024}]%
        {Paul_Zhu_Bayley_2024}
\bibfield{author}{\bibinfo{person}{Debalina~Ghosh Paul}, \bibinfo{person}{Hong Zhu}, {and} \bibinfo{person}{Ian Bayley}.} \bibinfo{year}{2024}\natexlab{}.
\newblock \showarticletitle{Benchmarks and Metrics for Evaluations of Code Generation: A Critical Review}.
\newblock  (\bibinfo{year}{2024}).
\newblock
\urldef\tempurl%
\url{https://doi.org/10.48550/ARXIV.2406.12655}
\showDOI{\tempurl}


\bibitem[\protect\citeauthoryear{Perdigão}{Perdigão}{[n.d.]}]%
        {perdigao_chatgpt_2023}
\bibfield{author}{\bibinfo{person}{Leone Perdigão}.} \bibinfo{year}{[n.d.]}\natexlab{}.
\newblock \bibinfo{title}{{ChatGPT}: a deep dive}.
\newblock
\newblock
\urldef\tempurl%
\url{https://leoneperdigao.medium.com/chatgpt-a-deep-dive-1feade9c4d77}
\showURL{%
\tempurl}


\bibitem[\protect\citeauthoryear{Ph.D}{Ph.D}{[n.d.]}]%
        {phd_decoder-only_2024}
\bibfield{author}{\bibinfo{person}{Cameron R.~Wolfe Ph.D}.} \bibinfo{year}{[n.d.]}\natexlab{}.
\newblock \bibinfo{title}{Decoder-Only Transformers: The Workhorse of Generative {LLMs}}.
\newblock
\newblock
\urldef\tempurl%
\url{https://cameronrwolfe.substack.com/p/decoder-only-transformers-the-workhorse}
\showURL{%
\tempurl}


\bibitem[\protect\citeauthoryear{Radford}{Radford}{[n.d.]}]%
        {Radford_2018}
\bibfield{author}{\bibinfo{person}{Narasimhan K. Salimans T. \& Sutskever~I. Radford, A.}} \bibinfo{year}{[n.d.]}\natexlab{}.
\newblock \showarticletitle{Improving language understanding by generative pre-training}.
\newblock  (\bibinfo{year}{[n.\,d.]}).
\newblock
\urldef\tempurl%
\url{https://www.mikecaptain.com/resources/pdf/GPT-1.pdf}
\showURL{%
\tempurl}


\bibitem[\protect\citeauthoryear{Ridnik}{Ridnik}{[n.d.]}]%
        {ridnik_state---art_2024}
\bibfield{author}{\bibinfo{person}{Tal Ridnik}.} \bibinfo{year}{[n.d.]}\natexlab{}.
\newblock \bibinfo{title}{State-of-the-art Code Generation with {AlphaCodium} - From Prompt Engineering to Flow Engineering}.
\newblock
\newblock
\urldef\tempurl%
\url{https://www.codium.ai/blog/alphacodium-state-of-the-art-code-generation-for-code-contests/}
\showURL{%
\tempurl}


\bibitem[\protect\citeauthoryear{Ridnik, Kredo, and Friedman}{Ridnik et~al\mbox{.}}{[n.d.]}]%
        {ridnik_code_2024}
\bibfield{author}{\bibinfo{person}{Tal Ridnik}, \bibinfo{person}{Dedy Kredo}, {and} \bibinfo{person}{Itamar Friedman}.} \bibinfo{year}{[n.d.]}\natexlab{}.
\newblock \bibinfo{title}{Code Generation with {AlphaCodium}: From Prompt Engineering to Flow Engineering}.
\newblock
\newblock
\urldef\tempurl%
\url{https://doi.org/10.48550/arXiv.2401.08500}
\showDOI{\tempurl}
\showeprint[arxiv]{2401.08500 [cs]}


\bibitem[\protect\citeauthoryear{Roumeliotis and Tselikas}{Roumeliotis and Tselikas}{[n.d.]}]%
        {roumeliotis_chatgpt_2023}
\bibfield{author}{\bibinfo{person}{Konstantinos~I. Roumeliotis} {and} \bibinfo{person}{Nikolaos~D. Tselikas}.} \bibinfo{year}{[n.d.]}\natexlab{}.
\newblock \showarticletitle{{ChatGPT} and Open-{AI} Models: A Preliminary Review}.
\newblock  \bibinfo{volume}{15}, \bibinfo{number}{6} (\bibinfo{year}{[n.\,d.]}), \bibinfo{pages}{192}.
\newblock
\showISSN{1999-5903}
\urldef\tempurl%
\url{https://doi.org/10.3390/fi15060192}
\showDOI{\tempurl}


\bibitem[\protect\citeauthoryear{Siroš, Singelée, and Preneel}{Siroš et~al\mbox{.}}{2024}]%
        {Siroš_Singelée_Preneel_2024}
\bibfield{author}{\bibinfo{person}{Ilja Siroš}, \bibinfo{person}{Dave Singelée}, {and} \bibinfo{person}{Bart Preneel}.} \bibinfo{year}{2024}\natexlab{}.
\newblock \showarticletitle{GitHub Copilot: the perfect Code compLeeter?}
\newblock  (\bibinfo{year}{2024}).
\newblock
\urldef\tempurl%
\url{https://doi.org/10.48550/ARXIV.2406.11326}
\showDOI{\tempurl}


\bibitem[\protect\citeauthoryear{Team, Anil, et~al\mbox{.}}{Team et~al\mbox{.}}{2023}]%
        {Gemini_Team_Anil_Borgeaud_Alayrac_Yu_Soricut_Schalkwyk_Dai_Hauth_Millican_et_al_2023}
\bibfield{author}{\bibinfo{person}{Gemini Team}, \bibinfo{person}{Rohan Anil}, {et~al\mbox{.}}} \bibinfo{year}{2023}\natexlab{}.
\newblock \showarticletitle{Gemini: A Family of Highly Capable Multimodal Models}.
\newblock  (\bibinfo{year}{2023}).
\newblock
\urldef\tempurl%
\url{https://doi.org/10.48550/ARXIV.2312.11805}
\showDOI{\tempurl}


\bibitem[\protect\citeauthoryear{Vincent}{Vincent}{[n.d.]}]%
        {vincent_deepmind_2022}
\bibfield{author}{\bibinfo{person}{James Vincent}.} \bibinfo{year}{[n.d.]}\natexlab{}.
\newblock \bibinfo{title}{{DeepMind} says its new {AI} coding engine is as good as an average human programmer}.
\newblock
\newblock
\urldef\tempurl%
\url{https://www.theverge.com/2022/2/2/22914085/alphacode-ai-coding-program-automatic-deepmind-codeforce}
\showURL{%
\tempurl}


\bibitem[\protect\citeauthoryear{Wang and Zhao}{Wang and Zhao}{[n.d.]}]%
        {wang_gemini_2023}
\bibfield{author}{\bibinfo{person}{Yuqing Wang} {and} \bibinfo{person}{Yun Zhao}.} \bibinfo{year}{[n.d.]}\natexlab{}.
\newblock \bibinfo{title}{Gemini in Reasoning: Unveiling Commonsense in Multimodal Large Language Models}.
\newblock
\newblock
\showeprint[arxiv]{2312.17661 [cs]}
\urldef\tempurl%
\url{http://arxiv.org/abs/2312.17661}
\showURL{%
\tempurl}


\bibitem[\protect\citeauthoryear{Wiggers}{Wiggers}{2023}]%
        {Wiggers_2023}
\bibfield{author}{\bibinfo{person}{Kyle Wiggers}.} \bibinfo{year}{2023}\natexlab{}.
\newblock \bibinfo{title}{Google unveils AlphaCode 2, powered by Gemini}.
\newblock
\newblock
\urldef\tempurl%
\url{https://techcrunch.com/2023/12/06/deepmind-unveils-alphacode-2-powered-by-gemini/}
\showURL{%
\tempurl}


\bibitem[\protect\citeauthoryear{Xiao and Zhu}{Xiao and Zhu}{[n.d.]}]%
        {xiao_introduction_2023}
\bibfield{author}{\bibinfo{person}{Tong Xiao} {and} \bibinfo{person}{Jingbo Zhu}.} \bibinfo{year}{[n.d.]}\natexlab{}.
\newblock \bibinfo{title}{Introduction to Transformers: an {NLP} Perspective}.
\newblock
\newblock
\showeprint[arxiv]{2311.17633 [cs]}
\urldef\tempurl%
\url{http://arxiv.org/abs/2311.17633}
\showURL{%
\tempurl}


\bibitem[\protect\citeauthoryear{Yan, Sha, Zhao, Li, Martinez‐Maldonado, Chen, Li, Jin, and Gašević}{Yan et~al\mbox{.}}{[n.d.]}]%
        {yan_practical_2024}
\bibfield{author}{\bibinfo{person}{Lixiang Yan}, \bibinfo{person}{Lele Sha}, \bibinfo{person}{Linxuan Zhao}, \bibinfo{person}{Yuheng Li}, \bibinfo{person}{Roberto Martinez‐Maldonado}, \bibinfo{person}{Guanliang Chen}, \bibinfo{person}{Xinyu Li}, \bibinfo{person}{Yueqiao Jin}, {and} \bibinfo{person}{Dragan Gašević}.} \bibinfo{year}{[n.d.]}\natexlab{}.
\newblock \showarticletitle{Practical and ethical challenges of large language models in education: A systematic scoping review}.
\newblock  \bibinfo{volume}{55}, \bibinfo{number}{1} (\bibinfo{year}{[n.\,d.]}), \bibinfo{pages}{90--112}.
\newblock
\showISSN{0007-1013, 1467-8535}
\urldef\tempurl%
\url{https://doi.org/10.1111/bjet.13370}
\showDOI{\tempurl}


\bibitem[\protect\citeauthoryear{Zayed, Mordido, Shabanian, Baldini, and Chandar}{Zayed et~al\mbox{.}}{2024}]%
        {Zayed_Mordido_Shabanian_Baldini_Chandar_2024}
\bibfield{author}{\bibinfo{person}{Abdelrahman Zayed}, \bibinfo{person}{Gonçalo Mordido}, \bibinfo{person}{Samira Shabanian}, \bibinfo{person}{Ioana Baldini}, {and} \bibinfo{person}{Sarath Chandar}.} \bibinfo{year}{2024}\natexlab{}.
\newblock \showarticletitle{Fairness-Aware Structured Pruning in Transformers}.
\newblock \bibinfo{journal}{\emph{Proceedings of the AAAI Conference on Artificial Intelligence}} \bibinfo{volume}{38}, \bibinfo{number}{20} (\bibinfo{date}{March} \bibinfo{year}{2024}), \bibinfo{pages}{22484–22492}.
\newblock
\urldef\tempurl%
\url{https://doi.org/10.1609/aaai.v38i20.30256}
\showDOI{\tempurl}


\bibitem[\protect\citeauthoryear{Zhang et~al\mbox{.}}{Zhang et~al\mbox{.}}{2024}]%
        {Zhang_Zhao_Liu_Zheng_Qi_Gu_Zhang_Dong_Tang_2024}
\bibfield{author}{\bibinfo{person}{Shudan Zhang} {et~al\mbox{.}}} \bibinfo{year}{2024}\natexlab{}.
\newblock \showarticletitle{NaturalCodeBench: Examining Coding Performance Mismatch on HumanEval and Natural User Prompts}.
\newblock  (\bibinfo{year}{2024}).
\newblock
\urldef\tempurl%
\url{https://doi.org/10.48550/ARXIV.2405.04520}
\showDOI{\tempurl}


\bibitem[\protect\citeauthoryear{Zhao}{Zhao}{[n.d.]}]%
        {zhao_github_2023}
\bibfield{author}{\bibinfo{person}{Shuyin Zhao}.} \bibinfo{year}{[n.d.]}\natexlab{}.
\newblock \bibinfo{title}{{GitHub} Copilot now has a better {AI} model and new capabilities}.
\newblock
\newblock
\urldef\tempurl%
\url{https://github.blog/2023-02-14-github-copilot-now-has-a-better-ai-model-and-new-capabilities/}
\showURL{%
\tempurl}


\end{thebibliography}

\end{document}